\documentclass[runningheads]{llncs}

\usepackage[T1]{fontenc}
\usepackage{graphicx}
\usepackage{amssymb}
\usepackage{colortbl}

\usepackage[hidelinks]{hyperref}

\begin{document}

\title{Explainability in Mechanism Design: Recent Advances and the Road Ahead}
\titlerunning{Explainability in Mechanism Design: Recent Advances and the Road Ahead}
\author{Sharadhi Alape Suryanarayana\inst{1,2} \and
David Sarne\inst{1} \and
Sarit Kraus\inst{1}}

\authorrunning{Suryanarayana et al.}
\institute{Department of Computer Science, Bar-Ilan University, Israel \and Centre for Ubiquitous Computing, University of Oulu, Finland\\ \email{sharadhi.as@gmail.com, david.sarne@biu.ac.il, sarit@cs.biu.ac.il}}

\maketitle              

\begin{abstract}
Designing and implementing explainable systems is seen as the next step towards increasing user trust in, acceptance of and reliance on Artificial Intelligence (AI) systems. While explaining choices made by black-box algorithms such as machine learning and deep learning has occupied most of the limelight, systems that attempt to explain decisions (even simple ones) in the context of social choice are steadily catching up. In this paper, we provide a comprehensive survey of explainability in mechanism design, a domain characterized by economically motivated agents and often having no single choice that maximizes all individual utility functions. We discuss the main properties and goals of explainability in mechanism design,  distinguishing them from those of Explainable AI in general. This discussion is followed by a thorough review of the challenges one may face when working on Explainable Mechanism Design and propose a few solution concepts to those.

\keywords{Explainability  \and Mechanism Design \and Justification.}
\end{abstract}

\section{Introduction}

Intelligent systems and automated decision-making are replacing and enhancing human decision-making nowadays to the extent that people are increasingly reliant on them \cite{mohseni2021multidisciplinary}. Despite the increased presence of such systems, people are not often aware that they are interacting with an AI-based system. Recognizing the need for transparency in this evolving policy and technology ecosystem, the ACM U.S. Public Policy Council (USACM) and ACM Europe Council Policy Committee (EUACM) codified a set of principles such as awareness, explainability, accountability, validation and testing to address this \cite{garfinkel2017toward}. Among which, \emph{Explainability}, which could be understood as a description in some form of the functioning of the system, has gained immense traction in the recent past.

Due to their opacity, domains with black-box algorithms like machine learning and deep learning have been extensively researched in the context of explainability. However, the need for explainability goes far beyond  black-box algorithms. For example in various multi-agent systems (MAS), where agents are self-interested, commonly arises a need to aggregate private preferences such as availability, budget constraints and geographical location of several agents into a collective decision in a socially desirable way.  Mechanism design, an important tool in economics and computer science, is one such research topic which is concerned with the development of a mechanism that takes into consideration the preferences of selfish and intelligent agents exhibiting strategic behavior while adhering to norms such as envy-freeness, budget-balancing and pareto-optimality \cite{procaccia2019axioms}. 
The applications of mechanism design can be found in various real-world and, in many cases, high-impact applications such as elections, rent division, resource allocation, and stable matching \cite{kominers2019good,procaccia2019axioms}.

Regardless of their extensive usage in the real-world, there is a renewed interest in designing and analyzing mechanisms to align with human values. This includes re-designing existing mechanisms to accommodate human preferences \cite{freedman2020adapting}, viewing existing practices for inclusive housing allocation from a game-theoretic perspective \cite{benabbou2018diversity}, empirical studies on human behavior \cite{tal2015study}, using the insights from empirical analyses to re-frame a mechanism \cite{obraztsova2017doodle}, and devising algorithms to justify the decision of a mechanism \cite{boixel2020automated}. Explaining the results to human participants is a natural and complementary extension to the pursuit of designing mechanisms that are more ``understandable" to humans.

Nevertheless, the road to Explainable Mechanism Design systems is replete with its own share of hurdles.
One key element in providing an explanation in such domains is the goal of the explanation and the measures of its success---whereas with a single user the system's goal is known, hence the explanation aims to improve her recognition of the optimality of the choice made, in settings of mechanism design, a user does not always know the system's goals since they may depend on other agents' preferences. 
This focus on preference aggregation of multiple agents, often associated with conflicting goals, may lead to a blatant compromise of the preferences of some of them. The explanations should therefore aim to increase user satisfaction by taking into account the system's decision, the user's and the other agents' preferences, the environment settings, and properties such as fairness, envy and privacy \cite{kraus2020ai}. 
In addition to the above intricacies, the presence of domains such as voting, scheduling and resource allocation in popular culture, without the necessity to be theoretically aware, has led to people forming their own irrational opinions which the explanations have to uproot \cite{d2020testing,mccune2019can,uhde2020fairness}.

We also note that even cases where social choice is merely a particular step in a multi-stage decision making process carried out by an agent making decisions on behalf of humans, call for Explainable Mechanism Design. This can hold even when the use of mechanism design is not explicit. Examples of such settings include algorithmic hiring and virtual democracy (an approach to automated decision making) which is used in autonomous vehicles and kidney exchanges to automate moral decisions, and recommendation systems to allocate food donations to recepient organizations \cite{freedman2020adapting,lee2019webuildai,noothigattu2018voting,schumann2020we}.

In this survey, we provide a comprehensive summary of the various threads of explainability in mechanism design. We note that while the broad theme of explainability translates to the same meaning with respect to both machine learning and mechanism design, there are a few differences between the premises of the two fields in terms of what leads up to generating explanations. In particular, we first provide a comparison between Explainable Mechanism Design and Explainable AI (XAI) with respect to the taxonomy, the purpose of explanations and who the explainees are. We then outline the methods of generating explanations in mechanism design. Finally, we elaborate on the challenges of conducting laboratory experiments on explainability in mechanism design and shed light on solution concepts combining insights from XAI and behavioral studies.

\section{Mechanism Design}

Various definitions for mechanism design have been suggested over the years \cite{nisan2001algorithmic,papadimitriou2001algorithms}.
Essentially, a mechanism can be seen as a ``communication system" where the participants send messages to each other and/or to a ``message center" and every collection of messages is assigned an outcome based on a pre-specified rule \cite{hurwicz1960optimality}. These messages are characterized by private information such as utility from an allocation (in rent division), preference over a set of candidates (in social choice theory), or willingness to pay for a good (in auctions). Thus, the mechanism is analogous to a machine that collects, processes and aggregates the private information of several agents in order to reach a desirable social outcome. In most cases the agents are self-interested and rational, and care only about maximizing their private utility with no guarantee that they will tell the truth. Therefore, a desired property of a mechanism is that the agents have no strategic incentive to deviate from truth-telling \cite{lavi2020mechanism}. A mechanism satisfying this property is considered \emph{incentive-compatible}, as every participating agent achieves the best outcome by reporting her true preferences \cite{lavi2020mechanism}.

Since incentive-compatibility is a desired feature, most mechanisms are designed to be incentive-compatible. Hence, even though truth-telling might fetch the least utility, say not winning the Vickrey-Clarke-Groves (VCG) auction \cite{lavi2020mechanism}, it is the best response for an agent in most mechanisms. However, as we observe in section \ref{sec:challenges}, due to repeated interactions, humans may be prone to their own biases. Explanations could be a tool to mitigate user biases as well. 

The two main branches of designing mechanisms are the axiomatic branch and the Bayesian branch \cite{carroll2019design}. In the axiomatic branch, the solution is supposed to satisfy a set of desired properties called axioms. Axioms are normative elements designed to conceptualize notions of reason such as fairness, justice and efficiency. 
Examples of such axioms include envy-freeness and pareto-optimality. In the Bayesian branch the solution achieves an optimal value for a given objective function such as expected revenue or projected loss.

\section{Motivations for Explainable Mechanism Design} \label{sec:motivation}
Consider the example of a rent division problem with 3 housemates, 3 rooms with a total rent of \$3. Housemate $i$ values room $i$ at \$3 and the other two rooms at \$0. One possible solution to this problem is to assign room 1 to housemate 1 at \$3 and rooms 2 and 3 to housemates 2 and 3 for free. Even though from an inter-personal perspective, this solution seems blatantly unfair to housemate 1, it is an envy-free solution, hence theoretically acceptable. Housemate 1 is indifferent between the three rooms, given their cost, while housemates 2 and 3 are overjoyed \cite{gal2016fairest}. While this is an overblown depiction of the problem with a rent division setting which has seen an immense improvement in the solutions proposed over the years \cite{gal2016fairest}, it is sufficient to illustrate the complexity with respect to devising explanations in mechanism design settings.

As explained in the former section, solutions in mechanism design are obtained by aggregating the preferences of several agents. The nature of the problem necessitates a ``social approach" where the solution is expected to aggregate said preferences in some acceptable manner. This can be achieved by mandating the solution to adhere to certain desirable criteria that are egalitarian in nature, maximizing a criterion of social welfare or using any other method that appreciates the social nature of the problem. The need to balance the preferences of several agents, which could be conflicting in nature might result in the solution not being in favor of a few of them. Multi-user Privacy Conflict due to varying privacy preferences of owners of shared content \cite{mosca2022explainable}, multi-attribute settings such as \emph{team formation} where the solution might not adhere to the preferences of all of the agents with respect to every attribute \cite{georgarabuilding} and settings such as the classical ``glove game" where the solution is non-intuitive yet theoretically sound \cite{nizri2022explainable} are other real-life examples of mechanism design that necessitate a nuanced approach to obtain the solution as well as to devise explanations.

In addition to the social nature of the problem, solutions in mechanism design face two hurdles. First, the issue of what is socially acceptable can vary according to perception, context and domain which can result in multiple solutions for the same problem. For example, in social choice theory, there are multiple voting rules due to the absence of a unique voting rule that satisfies Arrow's mandatory principles of fairness \cite{arrow1951social}. The second problem which is a consequence of the first, is that it is easy for the user to challenge the solution.
This requires the explanations to not only elaborate on how good the solution is but also how problematic another solution is. 

This is in sharp contrast with other domains such as machine learning, planning and recommender systems where a solution, good or bad, results from a definite and often complicated algorithm which needs to be broken down for user understanding \cite{chakraborti2020emerging,mohseni2021multidisciplinary,zhang2020explainable}. While solutions in other settings can also rely on metrics such as accuracy, the absence of a unique algorithm in mechanism design settings makes this problem hard as well. Hence, instead of arguing on the theoretical accuracy of the decision or how it results from a particular algorithm which is a common trait in the aforementioned domains, the focus of explanations in mechanism design should be on arguing how the decision is ``good" in its social context and how the preferences of the agents have led to the solution.

\section{Explainable Mechanism Design versus XAI}

While both Explainable Mechanism Design and XAI aim to explain certain decisions made by the system, in retrospect, there are several factors differentiating the two. Owing to the differences in the domains and the solution concepts, there is an obvious difference in the nature of explanations offered in Explainable Mechanism Design and XAI.  One way to reason about the differences in the nature of the explanation to be offered, is by considering the differences in the taxonomy to be used, the explainee and the goals of the explanations. Hence, we focus our comparison in these aspects. Since XAI has become a well established research area, whereas Explainable Mechanism Design is a newly emerging field, we use the first to lead the discussion, mapping and contrasting equivalent notions of Explainability in Mechanism Design accordingly.

\subsection{Taxonomy}

The common types of Explanations that are found in Explainability Studies are \emph{Explanations}, \emph{Justifications} and \emph{Interpretations}. With respect to mechanism design, the most relevant and consequently most researched capability is Justifiability. \emph{Justifications} deal with explaining the system's decision in terms of acceptable societal norms \cite{langley2019explainable}. These norms (i.e. axioms that formalize desirable social concepts such as fairness, justice and efficiency) are the foundation of many mechanisms.  Hence they are natural and, in many cases, effective contenders for explanations \cite{boixel2020automated,nizri2022explainable,procaccia2019axioms}. 

While there is no consensus on what constitutes an \emph{Explanation} in mechanism design, we adopt the idea proposed by Langley \cite{langley2019explainable}, who states that \emph{an intelligent system exhibits explainable agency if it can provide, on request, the reasons for its activities}. Still, the extent of research dedicated to this capability is somehow limited (see Table \ref{table:summarytable} later on). As for \emph{Interpretability}, this notion is completely absent from Explainable Mechanism Design.

For example, in the domain of fair division, explaining the decision by highlighting how the decision is envy-free is a \emph{Justification} \cite{lee2019procedural}. However, comparing the maximin solution (the solution that maximizes the minimum utility for every player thus resulting in the least disparity) to an arbitrary envy-free solution to demonstrate the superiority of the former solution amounts to an \emph{Explanation} \cite{gal2016fairest}.

However, in XAI, the relationship of Explanations and Justifications to the algorithm are reversed. Here, \emph{Explanations} act as an accurate proxy of the model while still being understandable to the human users \cite{arrieta2020explainable} and \emph{Justifications} defend the decision of the algorithm by explaining why it is a good one without necessarily focusing on how the decision was made \cite{rosenfeld2019explainability}. While nearly absent from \emph{Explainable Mechanism Design}, Interpretability which aims to enhance user understanding and comprehension of the model's decision-making process and predictions \cite{mohseni2021multidisciplinary} through \emph{Interpretations} \cite{rosenfeld2019explainability} is rigourously researched in XAI. It is also interesting to note that \emph{Justifiability} is the most researched capability in \emph{Explainable Mechanism Design} since there is a need to show how the decision abides by desirable social norms while the need to explain the functioning of black-box nature of the algorithms has led to \emph{Explainability} and \emph{Interpretability} enjoying the most attention from the XAI research community.

\subsection{Explanation Purpose}

Explanations may be provided for various purposes and goals, with partial equivalences between Explainable Mechanism Design and XAI.

\paragraph{Appreciating System Decision-Making.} In Explainable Mechanism Design, an appreciation of the system's decision translates to understanding how the preferences of the different agents are combined to obtain a collective decision \cite{boixel2020automated}. This is akin to understanding how different features contribute to the output in a machine learning model \cite{lai2019human}, which is the focus of XAI. Still, as explained in Section \ref{sec:motivation}, the task of increasing user appreciation of the system's decision making is more challenging in mechanism-design settings, as decisions need to be explained in their wider social context and are often plagued with impossibility theorems \cite{arrow1951social,roth1982economics}.

\paragraph{Improving User Trust and Reliability.} With the decision depending on the preferences of several agents, dissatisfaction is an inevitable evil in mechanism design which might affect the user's trust and reliability. The explanations presented  thus need to argue about the legitimacy of the decision and how, even if the decision is unfair to a particular agent, the mechanism as a whole has adhered to mandatory principles of fairness and the dissatisfied agent needs to make peace with it \cite{kraus2020ai}. In fact, as demonstrated in the work of Suryanarayana et al. \cite{Suryanarayana2022Justifying}, it is often the explanations provided to those participants for whom the winning candidate is the least preferred that are the most impactful. 
In XAI, unfavorable situations are also present (e.g., rejection of a loan application), however the prospect of improving the odds of the decision being in her favor through \emph{Counterfactual Explanations} exists \cite{mothilal2020explaining}.  This, unfortunately, cannot be achieved in \emph{Explainable Mechanism Design}.

\paragraph{Ensuring Accuracy of Decision-making.} In mechanism design, explanations can serve as a tool for the verification of results in order to ensure that the decision was made under a set of rules consistently applied in each setting \cite{belahcene2018accountable}. This somehow resembles
the use of XAI as a tool for bias mitigation and fairness assessment \cite{mohseni2021multidisciplinary} in cases where datasets are potentially biased and decision-making might be discriminatory.

\subsection{The Explainee}

The nature and mode of the explanation to be offered depend on who the explainee is and the purpose of providing explanations to her \cite{rosenfeld2019explainability}. One of the main recipients of explanations in \emph{Explainable Mechanism Design} are the end users or participants as they are the ones affected by the decision made.\footnote{In many cases the need to provide users with proper explanations is dictated by the regulator, e.g., in the case of GDPR guidelines \cite{kraus2020ai}.}
Unlike end users in XAI who are typically passive (as the system is making the decision for them, based on their data)  
users receiving explanations in mechanism design play an active role in the collective decision process taking place, by reporting their preferences. Examples of such explainees include a researcher whose grant proposal was rejected \cite{belahcene2018accountable}, one among many roommates who is assigned a particular room and rent based on her actively reported (and hopefully true) preferences \cite{gal2016fairest} or, a user in a hybrid domain like algorithmic hiring where a social choice function is applied in some part of the overall algorithm \cite{schumann2020we}.

Two classes of explainees in Explainable Mechanism Design enjoy some similarity with those of XAI. The first one is the \emph{Decision Maker} who does not need to be an expert in mechanism design but has to have relevant knowledge of the domain in order to make informed decisions. This could include the employees of a refugee resettlement agency who need to be able to override the decision proposed by the algorithm \cite{ahani2021dynamic} or the members of a committee who cannot decide on a specific voting rule and base the election on a set of desired properties (axioms) \cite{boixel2020automated}.
The Decision Makers are similar to \emph{Data Experts} in XAI who use explanations to visualize, inspect, tune and select models \cite{mohseni2021multidisciplinary}. The second class is that of an \emph{External Entity}, similar to its namesake in XAI \cite{rosenfeld2019explainability}, who is someone not directly interacting with the system, say an auditor who needs to ensure that the decisions made adhere to a set of rules and that there is no violation \cite{belahcene2018accountable}. In both classes of explainees, the requirements of the explanations to be produced in XAI and Explainable Mechanism Design overlap.

\section{Explanation Concepts}\label{sec:explanationconcepts}

As with XAI \cite{mohseni2021multidisciplinary}, Explainable AI Planning \cite{chakraborti2020emerging} and Explainable Recommendation \cite{zhang2020explainable}, we provide a brief overview of the theoretical aspects (natural contenders) and behavioral aspects (necessary for human comprehension) of the explanations available in literature. Table \ref{table:summarytable} depicts a breakdown of the surveyed papers with respect to the different concepts and the evaluation methods discussed in the next section. 

 \begin{table*}[t]
  \caption{Surveyed literature organized by Explanation Concepts -- Normative Characterization (NO), Attributive (AT), Contrastive (CO), Argumentative (AR), Visualization (VI) and Evaluation Methods -- Theoretical Properties (TP), Computational Complexity (CC), Empirical Analysis (EA) and User Studies (US).}
  
  \begin{tabular}{|p{3.5cm}|p{2.5cm}|p{0.7cm}p{0.7cm}p{0.7cm}p{0.7cm}p{0.7cm}p{0.7cm}p{0.7cm}p{0.7cm}p{0.7cm}|}
\hline
 \multicolumn{11}{|c|}{Explanation Concepts and Evaluation Methods} \\
 \hline
 
                        Work & Setting & NO & AT & CO & AR & VI & TP & CC & EA & US  \\ \hline
                        Ahani et al. \cite{ahani2021dynamic} & Refugee Resettlement &  &\checkmark &\checkmark & & \checkmark & & & & \\ 
    \rowcolor[gray]{.9} Belahcene et al. \cite{belahcene2018accountable} & Approval Sorting &\checkmark & & \checkmark & \checkmark& & \checkmark & & &\\ 
                        Boixel and Endriss \cite{boixel2020automated} & Voting & \checkmark & & & & & \checkmark & & \checkmark & \\ 
    \rowcolor[gray]{.9} Boixel et al. \cite{boixel2022calculus} & Voting & \checkmark & & & & & \checkmark & & & \\
                        Boixel and de Haan \cite{boixel2021complexity} & Voting & \checkmark & & & & & & \checkmark & & \\ 
    \rowcolor[gray]{.9} Cailloux and Endriss \cite{cailloux2016arguing} & Voting & \checkmark & & & \checkmark & &\checkmark & & & \\ 
                        Gal et al. \cite{gal2016fairest} & Rent Division & \checkmark & \checkmark & \checkmark & & \checkmark & & & &  \checkmark \\
    \rowcolor[gray]{.9} Georgara et al. \cite{georgarabuilding} & Team Formation & & \checkmark & \checkmark & & & & & \checkmark & \\
                        Kirsten and Cailloux \cite{kirsten2018towards} & Voting & \checkmark & & \checkmark &  \checkmark & & & & \checkmark &  \\
    \rowcolor[gray]{.9} Knapp \cite{knapp2022justification} & Matching Theory & \checkmark & & & & & \checkmark & & \checkmark & \\       
                        Lee et al. \cite{lee2019procedural} & Rent Division & \checkmark & \checkmark & \checkmark & & \checkmark & & & & \checkmark \\ 
    \rowcolor[gray]{.9} Mosca and Such \cite{mosca2022explainable} & Multi-User Privacy Conflict & \checkmark & \checkmark & \checkmark & \checkmark & & & & & \checkmark \\
                        Nardi et al. \cite{NardiEtAlAAMAS2022} & Voting & \checkmark & & & & &  \checkmark & & \checkmark & \\
    \rowcolor[gray]{.9} Nizri et al. \cite{nizri2022explainable} & Payoff Allocation & \checkmark &  & & & & & & & \checkmark \\
                        Peters et al. \cite{peters2021market} & Voting & & \checkmark & & & & \checkmark & \checkmark & \checkmark & \\
    \rowcolor[gray]{.9} Peters et al. \cite{peters2020explainable} & Voting & \checkmark & & & & & \checkmark & & \checkmark & \\
                        Pozanco et al. \cite{Pozanco2022Explaining} & Scheduling & & \checkmark & \checkmark & & & & & & \checkmark \\
    \rowcolor[gray]{.9} Suryanarayana et al. \cite{Suryanarayana2022Justifying} & Voting & \checkmark & & \checkmark & & & & & & \checkmark \\
                        Zahedi et al. \cite{zahedi2020didn} & Task Allocation& & \checkmark & \checkmark & \checkmark & & \checkmark & & & \checkmark \\

  \hline    

 \end{tabular}
 \label{table:summarytable}
 \end{table*}

\subsection{Norms versus Attributes}
As mentioned earlier, mechanism design has two defining characteristics - the private information (such as preference and cost) of the agents participating in it and the requirement for the solution to recognize its social nature. Both of these can be used to devise explanations. Norms that formalize the desirable social traits are the foundation of solutions in mechanism design and can hence be used to extol its virtues. Attributes, on the other hand, quantify the stake a given agent has in the mechanism. Explanations that relate the solution to an agent's individual stakes can be effective in helping her appreciate the impact of the solution from a selfish perspective and thus convince her. For example, convincing a housemate in a rent division setting that the decision is envy-free amounts to a normatively characterized explanation \cite{lee2019procedural} while the comparison of the maximin (the solution that maximizes the minimum utility for every player thus resulting in the least disparity) solution to an arbitrary envy-free solution to demonstrate the lower disparity achieved by the former solution is an attributive explanation \cite{gal2016fairest}. In the following paragraphs, we elaborate on diverse settings where both norms and attributes have been used to devise explanations.

\paragraph{Normative Characterization.}

Formally in mechanism design, axioms are used to capture the social norms that the solution is expected to adhere to. 
Procaccia \cite{procaccia2019axioms} advocates for the use of axioms to not only be used for designing a mechanism but also to justify its solutions with an example of the not-for-profit website \emph{Spliddit} \cite{goldman2015spliddit}. Justifying an outcome using a set of agreed upon axioms, without having to depend on a particular rule has found a special appeal in the domain of social choice theory, where no unique outcome can be obtained while following fair voting procedures \cite{arrow1951social}.

In  social choice theory, given a voting profile, Cailloux and Endriss \cite{cailloux2016arguing} developed a logic-based language to construct arguments for and against specific outcomes. Using the elements of the proposed language, an algorithm to justify the Borda outcome given a voting profile was developed. Building on said approach, Boixel and Endriss \cite{boixel2020automated} developed a formal notion of justification based on the definition of Langley \cite{langley2019explainable} and an algorithm based on constraint programming to compute the justifications using any set of axioms. To counter the computational complexity of the aforementioned algorithm \cite{boixel2021complexity}, Nardi et al. \cite{NardiEtAlAAMAS2022} used a combination of instance graphs and state-of-the-art SAT solvers to design an algorithm that can provide viable justifications. To enhance the readability of the justifications using the axiomatic approach, Boixel et al. \cite{boixel2022calculus} developed a tableau-based calculus. Using a combination of SAT solving and Answer Set Programming to implement the calculus, the authors provide an insight into how the justifications look.  

The evolution from Boixel and Endriss \cite{boixel2020automated} to Boixel et al. \cite{boixel2022calculus} helps visualize the transformation from a non-automated procedure to an automated procedure, from unstructured justifications to structured justifications and, from manual post processing to obtain the justifications to tableau-based rendering of the justifications for enhanced readability. A demonstration summarizing the application of the aforementioned techniques proposed \cite{boixel2020automated,boixel2022calculus,nardi2021graph,NardiEtAlAAMAS2022} to find justifications given a normative basis can be found in Boixel et al. \cite{BoixelEtAlIJCAI2022}. 
Furthermore, the approach used by Boixel and Endriss \cite{boixel2020automated} has been extended to matching theory \cite{knapp2022justification} where an algorithm is designed to justify outcomes that are of interest to a given agent (local outcomes) instead of the whole outcome.

The axiomatic characterization is also used to derive justifications for the results of approval voting \cite{peters2020explainable} and non-compensatory approval sorting \cite{belahcene2018accountable}.   In the broader sense, non-compensatory approval sorting and voting are concerned with aggregating collective information into a single decision. The reviewed literature on justifying the results of a voting mechanism reveal all of the preference ballots. However, Belahcene et al. \cite{belahcene2018accountable} show that the classification based on the binary judgements of the participants is compliant with the decision making process by revealing minimal information that is backed by theoretical properties.

One of the key elements used in some of the papers is Automated Reasoning (AR) using SAT or SMT solvers \cite{belahcene2018accountable,boixel2022calculus,nardi2021graph}. This combination of AR with social choice theory can be used to identify if a voting rule satisfies a particular axiom (thus arguing against it) \cite{kirsten2018towards} as well as verifying the correctness of the system output \cite{boixel2021complexity}. 

User studies to test axiom-based explanations were also successful in increasing satisfaction. Suryanarayana et al. \cite{Suryanarayana2022Justifying} tested explanations based on features constructed from axioms in the domain of ranked-choice voting while Nizri et al. \cite{nizri2022explainable} used the axiomatic characterization of Shapley value \cite{shapley1953value} to come up with explanations in the domain of fair division. The research carried out by Nizri et al. \cite{nizri2022explainable} is significant in two ways. First, the solution that is being justified, i.e. Shapley value, satisfies all of the desired properties of fair division \cite{hart1989shapley}. Second, the axiomatic characterization is not only used to justify the solution but also to come up with the algorithm to generate explanations. The authors decompose the coalitional game into sub-games and generate explanations for each of these games based on the additivity axiom which states that the sum of Shapley allocations in each sub-game is equal to the Shapley allocation in the original game. The explanations were successful in convincing the participants that the allocation was fair.

An exception in terms of the norms used can be found in the work of Mosca and Such \cite{mosca2022explainable} in the domain of Multiuser Privacy who propose an explainable agent called ELVIRA that collaborates with other ELVIRA agents to identify the optimal sharing policy for shared content. Here, instead of axioms, the authors use a socio-cultural theory of human values by Schwartz \cite{schwartz2012overview} known as the \emph{theory of basic values}. The explanations however, are based on both values and individual attributes, i.e. the privacy preferences of the participants.

\paragraph{Attributive Explanation.}
If norms capture societal acceptance, attributes quantify personal interests. Hence, explanations that relate to the individual attributes of the participants have also been fruitful in increasing participant satisfaction

Zahedi et al. \cite{zahedi2020didn} compare the \emph{cost} of a proposed allocation to the cost of the counterfactual allocation proposed by the participant. Ahani et al. \cite{ahani2021dynamic} depict the change in \emph{employment score} if the refugee allocation proposed by the algorithm needs to be changed. 
A tangential direction termed \emph{priceability} where voters spend money on buying candidates which forms an intuitive explanation for the committee selected in approval-based committee elections was proposed by Peters et al. \cite{peters2021market}. Explaining outcomes based on individual attributes enables the comparison of solutions that are equally good in terms of theoretical requirements. For example, Gal et al. \cite{gal2016fairest} explain their optimal rent division solution by comparing it to another envy-free rent division. 

Several explanation generation methods are procedure-agnostic i.e., do not focus on the procedure that leads to the outcome. Here, a framework is developed to encode the different facets of the problems such as explanations, queries and constraints. Notable examples include encoding the Justification Generation Problem for collective decisions into a Constraint Network \cite{boixel2020automated}, developing a generic procedure for providing justifications for Team Formation Algorithm (TFA) while keeping the TFA intact \cite{georgarabuilding} and using Mixed-Integer Linear Programming (MILP) to explain why the preferences of a participants were not satisfied \cite{Pozanco2022Explaining} while designing a preference-driven schedule. While using such frameworks simplifies the process of finding an explanation, adequate care needs to be taken to convert the explanations provided by the system into a readable form. One solution to this problem is using explanation templates \cite{georgarabuilding,Pozanco2022Explaining}.

\subsection{Catering to the Human Mind}

Understanding how human-beings explain and respond to explanations can reveal important insights into how explanations of a system can be presented. Two such behavioral modes of explanations that are considered effective and that have found applications in mechanism design are \emph{Contrastive} explanations and \emph{Argumentative} explanations \cite{miller2019explanation}. Given the fact that mechanism design settings are social in nature, it is imperative that the behavioral nature of the explanations are attended to.  
Most of the papers that use a behavioral element in their explanations incorporate an element of social-interaction \cite{miller2019explanation}, which is necessary for a layman to comprehend the functioning of a complicated AI-based system.

As mentioned earlier, mechanism design settings suffer from the issues of familiarity, non-uniqueness of the solution and the solution not being in favor of the participants. This provides the perfect ground for the participants to challenge the solution. Identifying this, there is a great deal of interest in devising \emph{contrastive} explanations that provide reasons for why a particular event did not occur as opposed to why a particular event did  \cite{miller2019explanation}. From Table \ref{table:summarytable}, we can see that in nearly all of the cases where the explanations are based on the individual attributes of the agents, they are contrastive. In these scenarios, contrastive explanations can help the participant compare the difference in her utility across different solutions and thus appreciate the decision better.

In Suryanarayana et al \cite{Suryanarayana2022Justifying}, a contrastive explanation comparing the winning candidate to the participant's most preferred candidate were found to increase user satisfaction and acceptance the most when the winning candidate was the least preferred option of the participant. Similarly, in Mosca and Such \cite{mosca2022explainable}, contrastive explanations were found to be more appealing than general descriptive explanations when the recommended solution was different from the participant's preference.
Contrastive explanations are also especially useful in multi-attribute/multi-preference settings where the outcome may not align with all of the preferences of any participant. Other notable studies that uses the contrastive approach are of Georgara et al. \cite{georgarabuilding} that provide explanations for both collaboration queries (questioning team formation) and assignment queries (challenging the assignment of teams and individuals to tasks) at individual, local and global levels, and the work of Pozanco et al. \cite{Pozanco2022Explaining} which provides contrastive explanations regarding the unsatisfied preferences of the participants while ensuring that the explanation is relevant to the participant.    

As far as argumentation is concerned, the presence of umpteen conflicting axioms is an encouraging premise to build an argumentation framework. This is demonstrated by Cailloux and Endriss \cite{cailloux2016arguing} who developed a formal framework for presenting arguments favoring a particular outcome. Zahedi et al. \cite{zahedi2020didn} present the case for the a suggested task allocation by demonstrating how a negotiation based on the counterfactual task allocation proposed by the participant can lead to a higher cost. 
Mosca and Such \cite{mosca2022explainable} base their explanation on the argumentation scheme used to obtain the optimal solution. Both in Zahedi et al. \cite{zahedi2020didn} and Mosca and Such \cite{mosca2022explainable} argumentation is used for devising the optimal solution which was then organically extended to generating explanations in favor of the outcome.

\emph{Visualization} is a tool that is often viewed as a less technical means of conveying complex theoretical concepts \cite{barwise1991visual}. Human-in-the-loop systems are a natural extension to mechanism design that caters to capturing reality better. Here, the algorithms have the capacity to process large volumes of data while expert insights are required to handle the inherent uncertainty of the real world. Hence, in addition to enabling easier comparisons \cite{gal2016fairest}, visual tools can also be used to support human decision-making. 

Notable illustrations of visualization can be found in the case of the resettlement agency \emph{HIAS} that is involved in resettlement of refugees into communities in the USA. The matching software \emph{Annie}\textsuperscript{TM} \emph{MOORE} enables the employees to override the proposed allocation by revealing the updated statistics so that no change will have a grievous impact \cite{ahani2021dynamic}. 

Explaining the outcome through effective visualizations can aid in enhancing the appreciation of fairness, a theoretical notion that is the bread and butter of mechanism design, as was observed by Gal et al. \cite{gal2016fairest}. While Ahani et al. \cite{ahani2021dynamic} capture practical elements such as indivisible families of refugees, batching and, an unknown number of refugee arrivals in the context of refugee resettlement, Gal et al. \cite{gal2016fairest} provide the \emph{fairest} division of rent subject to envy-freeness. In both of these cases, the practical relevance and efficacy of the proposed algorithm is demonstrated with the help of explanations. Hence, user studies with explanations can be seen as a complementary extension to establishing the superiority of novel algorithms while comparing them to existing state-of-the-art algorithms.
It is also interesting to note that  
users of the website \emph{Spliddit}, the platform from which data was used by Gal et al. \cite{gal2016fairest}, are provided with a detailed explanation on why the proposed rent division is fair, thus signifying the utility of explanations in everyday usage. Lee et al. \cite{lee2019procedural} used visualizations to both provide an elaborate breakdown of the process as well as let the participant experiment with different values in the website \emph{Spliddit}, to observe the changes. When the participants were shown only their outcome, they perceived the results as unfair while when they were shown the preferences and outcomes for all of the participants in the group, the participants perceived the result as fair.

From Table \ref{table:summarytable}  
it can be observed that Normative Characterization and Contrastive Explanations are extensively used in comparison to their other theoretical and behavioral counterparts, respectively.

\section{Evaluation Methods}

There are several dimensions for evaluating methods of Explainable Mechanism Design from theoretical as well as practical perspectives. We provide a description on each of them in the following paragraphs.

\paragraph{Theoretical Properties.} Building explanations on the foundation of concepts like Axiomatic Characterization, Logic-based Programming and Automated Reasoning necessitates these methods to be supported by rigid theoretical norms. Examples include proof of an explanation given the problem instance \cite{cailloux2016arguing,zahedi2020didn}, uniqueness of the outcome and justification given a voting profile and normative basis \cite{boixel2020automated}, the correctness of a tableau-based calculus for generating explanations \cite{boixel2022calculus} and an upper bound on the number of steps required to generate justifications to ensure readability \cite{peters2020explainable}.

\paragraph{Computational Complexity.} Practical feasibility of any explanation-generation method is tantamount to its real-life application. Very few authors have addressed this aspect in their work. Exceptional examples include, Peters et al. \cite{peters2021market} that demonstrate the polynomial-time verifiability of their proposed heuristic algorithm. Boixel and de Haan \cite{boixel2021complexity} prove the intractability of finding and generating justifications given a normative basis.

\paragraph{Empirical Analysis.} Empirical insights act as an intermediate between theoretical guarantees and experimental results. Running the explainability studies on real or synthetic datasets can help compare the running times of several explanation generation methods and pick the fastest one \cite{nardi2021graph}, help understand the step-by-step breakdown of the explanation generation method \cite{peters2020explainable}, disclose interesting insights about different statistical cultures (e.g., probability distribution of election profiles) that might help in the development of personalized explanations \cite{nardi2021graph}, and identify and evaluate metrics for the evaluation of explanation-generation methods before deploying them in real-time studies \cite{georgarabuilding}.

\paragraph{User Studies.} 
User studies are an effective means for examining the consequences of explanations on aspects such as reliability, satisfaction, trust and conviction.
While it is always desirable to conduct the experiment with the actual participants of a mechanism as in the case of \emph{Spliddit} \cite{gal2016fairest}, experiments conducted with synthetic data using platforms such as Amazon Mechanical Turk (AMT) or laboratory settings are a great starting point \cite{nizri2022explainable,Suryanarayana2022Justifying}. 

In addition to the obvious purpose of helping determine the impact of explanations, user studies can also be used to provide insights on curating effective explanations. For example, Suryanarayana et al. \cite{Suryanarayana2022Justifying} hinted at a user bias in favor of plurality voting rule while Mosca and Such \cite{mosca2022explainable} used the experimental insights to devise a hybrid explanation framework and improve the wording of the explanations.

From Table \ref{table:summarytable} we can observe that there is a good mix of all of the evaluation methods in the literature. However, performing \emph{Empirical Evaluation} and \emph{User Studies} to evaluate explanation-generation methods in mechanism design is challenging and we address this issue in detail in the next section.

\section{Challenges and Possible Solutions}\label{sec:challenges}
Despite the rapid increase in interest in explainable systems for mechanism design, the progress made in this field is still far behind compared to XAI. One of the main challenges is that of testing. As far as testing for the efficacy of explanations is considered, the ideal premise would be testing with real users, as in \emph{Spliddit} \cite{gal2016fairest} (where  explainees are the actual renters in a rent division setting) or in the work of Pozanco et al. \cite{Pozanco2022Explaining} (where a real restrictive \emph{return to office} scenario due to the COVID-19 pandemic was tackled). 
However, the development of such evaluations is expensive and users of such real-life settings are often inaccessible to the research community. We therefore outline a few challenges in designing nearly realistic experiments that can be conducted in laboratories or platforms like AMT.

\paragraph{User Behavior.} Human participants in a mechanism are poles apart from the perfect agents modeled in theory and exhibit short-sighted and downright irrational behavior. Instances of such behavior include reward divisions that adhere to weaker axioms than those that characterize Shapley value \cite{d2020testing}, playing dominated strategies in cooperative settings such as fair division and bargaining \cite{10.1145/3328526.3329592}, and performing manipulations that can be captured by simple heuristics \cite{mennle2015power}. 
Any \emph{social} explanation \cite{miller2019explanation} thus catering to the expected selfish interests, while the participants betray the same, may defeat the purpose of explanations. For example, in an experiment conducted on human behavior in voting, Tal et al. \cite{tal2015study} report that the participants exhibit herding by disregarding their most preferred candidate and voting for the candidate with the most first place votes in a predictive poll, even though it is the least optimal choice for them. In this case, framing a contrastive explanation comparing the participant's most preferred candidate and the winning candidate (which might be the candidate the participant voted for) would be counter-productive.

 A natural solution to the problem of mismatched behavior is building predictive models using behavioral, game-theoretic and machine learning tools. Examples include models for predicting human decisions in plurality voting \cite{fairstein2019modeling}, approval voting \cite{scheuerman2021modeling} and auctions \cite{noti2021bid}. The benefits of such models are twofold. First, it might help the explanations to be more \emph{selective} \cite{miller2019explanation} by shedding light on what is important to the explainee. For example, in the game-theoretic model of human behavior in Doodle Polls \cite{obraztsova2017doodle}, the concept of \emph{Social Bonus} is proposed in order to reason why voters vote for unpopular slots. 
Consequently, contrastive explanations comparing the winning candidates to the unpopular ones can be discarded as the latter are insignificant to the voter. 

The second and rather consequential utility from such models is that they might help identify the sub-optimal manipulations of the participants which can be contrasted with the optimal choice. In that context, an interesting hypothesis to investigate is if and what kind of explanations can bring irrational humans closer to the rational agents modeled in literature. This will open new avenues for \emph{Interpretability} in mechanism design which has not received as much attention as in the XAI literature \cite{rosenfeld2019explainability}.

\paragraph{User Biases.}  The prevelance of mechanisms in society has led to humans forming their own prejudices such as favoritism for plurality voting rule \cite{mccune2019can,peters2020explainable}, altruism towards non-performing participants \cite{d2020testing} and a disdain towards algorithmic decisions as being far from reality \cite{lee2017algorithmic,mckelvey1992experimental,uhde2020fairness}.

Long before building explainable systems was considered, researchers invested efforts into manually explaining technical jargon to non-expert participants. Notable examples include acquainting participants of a centipede game with backward induction \cite{mckelvey1992experimental} and measuring the frequency of violation of fairness criteria in voting \cite{mccune2019can}. Coupling these ideas with biases such as \emph{automation bias}, where a user believes that a computing system is more knowledgeable and ``intelligent" than it is, is a direction worth exploring \cite{goddard2012automation}. In addition, comparing different modes of presenting explanations such as textual and visual, both of which are extensively used in Explainable Recommender Systems \cite{zhang2020explainable}, can strengthen the efficacy of explanations.  It is also noteworthy that human intelligence can be leveraged to not only rate explanations but also to provide explanations, thus providing valuable insights into human factors that might be useful for generating convincing explanations \cite{Pozanco2022Explaining,Suryanarayana2022Justifying} .

\paragraph{Lack of Data.} A useful tool in bridging the gap between idealized agent behavior and flawed human behavior is empirical analysis and subsequent modeling of human participants in the different mechanism design settings. Also, as mentioned earlier, empirical analysis can act as an intermediate stage between theoretical guarantees and user studies while revealing interesting insights.

However, there are not many datasets in the domains of Computational Social Choice and Preference Reasoning publicly available \cite{mattei2021closing}. Also, while collecting, preserving and presenting data on private preferences, adequate care needs to be taken to ensure that user privacy is preserved \cite{kraus2020ai}.

Naturally, the obvious solution to the lack of data is to develop tools for efficient data collection. One such very useful collection of datasets in the domain of Computational Social Choice is \emph{Preflib}\footnote{\url{https://www.preflib.org/}} which was used by Nardi \cite{nardi2021graph} to examine the practical utility of the proposed algorithm. However the process of data collection is easier said than done. Replicating real-life settings in order to get people to report their preferences, even manipulated ones, is a mammoth task. Anonymizing data is an effective way to protect the privacy of the participants. An alternate technique to preserve privacy was used by Gal et al. \cite{gal2016fairest} where the original valuations for the rooms were perturbed by an acceptable margin and presented to the explainees.

\paragraph{Simulating Synthetic Environments.} In order to evaluate the efficacy of explanations, it is vital to have the participant interested in the explanations. In mechanism design, these interests are captured by notions such as preferences, utility and costs which are easy to conceptualize but difficult to replicate and/or induce in lab experiments. XAI, even though tasked with explaining complex algorithms, enjoys relatability with experiments such as image classification \cite{nourani2019effects}, review classification \cite{lai2019human} and selection of a competent agent \cite{amir2019summarizing}. This enables the design of interactive experiments where explanations can be sneaked in without being explicit, hence eliciting an organic response from the participant.

Inspired by the experimental design in XAI, gamification of the problems is a good starting point. Tailoring games such as the centipede game for bargaining \cite{mckelvey1992experimental} and share-the-loot game for resource allocation \cite{d2020testing} to accommodate explanations and with the reward tied to the performance of the participant is an idea worth testing. The presence of a monetary reward inadvertently engrosses the participant, thus eliciting a realistic response. Some other tested methods of invoking user interest in synthetic lab experiments were done using the concept of bonus from the winning candidate in ranked choice voting \cite{Suryanarayana2022Justifying} and asking the participant to imagine themselves in the setting and extracting their preferences through meticulously designed questionnaires \cite{mosca2022explainable}.

Another way to stimulate the interest of participants in explanations might be to leverage the diversity of axioms to build argumentation systems augmented with human input on how convincing the arguments are \cite{rosenfeld2016providing}. The conversion from passive listeners of explanations to active debaters of arguments might trigger a passionate yet honest response from the participants.

In addition to the above ideas, tools used in Social Psychology such as \emph{Experimental Vignette Methodology (EVM)} \cite{aguinis2014best} and online testing methods like A/B testing used in Explainable Recommendation \cite{zhang2020explainable} offer valid premises for developing tests for Explainable Mechanism Design.

\section{Conclusion}

In this paper, we survey explainability in mechanism design, provide an overall picture of the various concepts around it and shed light on the challenges faced by researchers in the domain.

While we do propose several workarounds to overcome the aforementioned challenges, we emphasize that implementing each of these is a non-trivial task per se and calls for collaborations between researchers in mechanism design, human-agent interaction, software engineering, and psychology. We hope that both experienced as well as budding researchers find this survey helpful in designing and improving explainability in mechanism design. We also envision a future where designing mechanisms aligned with human values and Explainable Mechanism Design complement each other.

\section*{Acknowledgements} This work was supported in part by the Data Science Institute at Bar-Ilan University, the EU Project TAILOR under grant 952215 and the Israeli Ministry of Science \& Technology under grant 89583. The research was carried out with the technological support and funding from the HRI Consortium -- the Israel Innovation Authority. Sharadhi Alape Suryanarayana is grateful for the President's Scholarship and Erasmus+ Global Mobility Programme that has supported this research. 

%
%
%
 \bibliographystyle{splncs04}
 \bibliography{samplepaper}
%





\end{document}